\documentclass{aa}
\usepackage{graphics}
\begin{document}

   \thesaurus{08     
              (09.08.1;  
               09.09.1;  
               09.19.2;  
               13.25.5)} 
   \title{The surroundings of the superluminal source GRS~1915+105}

   \author{L. F. Rodr\'\i guez\inst{1}
          \and
          I. F. Mirabel\inst{2,3}
          }

   \offprints{L. F. Rodr\'\i guez}

   \institute{Instituto de Astronom\'{\i}a, UNAM, 
              J. J. Tablada 1006, Morelia, Michoac\'an, 58090 M\'exico\\
              email: luisfr@astrosmo.unam.mx
         \and
             CEA/DSM/DAPNIA/Service D'Astrophysique, Centre d'Etudes de Saclay,  
             F-91191 Gif-sur-Yvette, France\\ 
             email: mirabel@discovery.saclay.cea.fr
         \and 
             Instituto de Astronom\'\i a
             y F\'\i sica del Espacio, C.C. 67, Suc. 28. 1428, 
             Buenos Aires, Argentina\\
             }

   \date{Received ; accepted }

\titlerunning{Surroundings of GRS 1915+105}
\authorrunning{Rodr\'\i guez \& Mirabel} 

   \maketitle

   \begin{abstract}

We have carried out radio studies of the surroundings
of the superluminal microquasar GRS~1915+105. Our main
goal was to understand the possible relation of GRS~1915+105
with two infrared/radio sources that appear symmetrically
located with respect to GRS~1915+105 and aligned with
the position angle of the relativistic ejecta.
We have also studied a nearby supernova remnant 
to test if the event that created the remnant could have been
the progenitor of
this hard X-ray binary.

   \end{abstract}

%

\section{Introduction}

Great advances have been made at radio, infrared, and X-ray wavelengths in
the study 
of the microquasar GRS~1915+105 (Mirabel \& Rodr\'\i guez 1994; 1998),
the first superluminal source found in our Galaxy. 
Much less attention has
been paid to its surroundings, in particular trying to 
search for nearby supernova remnants that
could be associated to the formation of this hard X-ray binary,
and also to understand where
in the interstellar medium
is the large kinetic energy of the relativistic ejecta 
(reaching 10$^{43}$ ergs; Rodr\'\i guez \& Mirabel 1999) being dissipated.

In this paper we present a large-scale mosaic image at 20-cm continuum of the environment of
GRS~1915+105 searching for nearby supernova remnants
and other extended objects.
We also report on extensive radio continuum and recombination line
studies of two infrared/radio sources,
IRAS~19124+1106 and IRAS~19132+1035, that appear symmetrically
located with respect to GRS~1915+105 and aligned with
the position angle of the relativistic ejecta.
All the observations were made with the Very Large Array of 
NRAO\footnote{The National 
Radio Astronomy Observatory is operated by
Associated Universities, Inc., under cooperative agreement with the
USA National Science Foundation.}.

\section{Large scale 20-cm continuum map}

   \begin{figure}
\resizebox{\hsize}{!}{\includegraphics{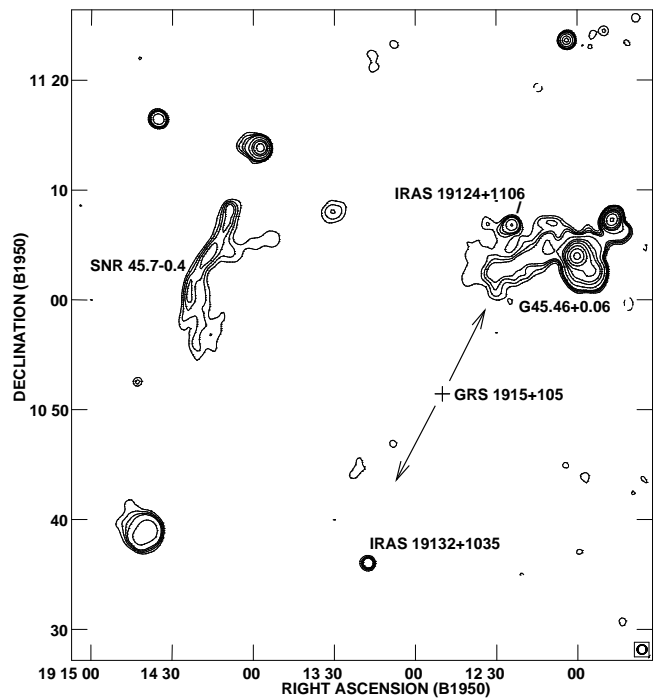}}
     \caption{VLA mosaic at 20-cm of the surroundings of GRS~1915+105.
Contours are -4, 4, 6,
8, 10, 20, 40, 60, 100, 200, 400, and 800 times 3 mJy beam$^{-1}$.
The half power contour of the beam is shown in the bottom right corner.
The arrows indicate the position angle of the relativistic ejecta.}
         \label{Fig1}%
    \end{figure}
During 1997 December 14 we did a 3$\times$3 mosaic of the surroundings of GRS~1915+105
at 20-cm continuum using the VLA in the D configuration. The resulting image is shown in 
Figure 1.
The most important objects in the region are identified in this Figure.
GRS~1915+105 was undetectable at the epoch of the observations and its position
is marked with a cross.
To its NW and SE we can see the two IRAS sources,
IRAS~19124+1106 and IRAS~19132+1035, that will be discussed below.
IRAS~19124+1106 appears projected in close association with a complex of H~II regions
that includes well-studied sources as G45.46+0.06.

To the NE of GRS~1915+105 we can see the supernova remnant SNR 45.7-0.4.
This remnant was first observed at 30.9 MHz by Kassim (1988) with the 
Clark Lake telescope. It was later observed by F\"urst et al. (1987) at 
1.4, 2.7, 4.7, and 10.6 GHz with the Bonn telescope. The Bonn 
observations show spectral indices in the range of -0.3 to -0.5, and at 
4.75 GHz 21\% polarization in the arc-shaped structure shown in Figure 1. 

Are SNR 45.7-0.4 and GRS~1915+105 related? In the sky, and assuming a distance of 12.5 kpc,
GRS~1915 appears displaced by about 40 pc from the centroid of SNR 45.7-0.4.
The proper motion study of GRS~1915+105 by Dhawan, Mirabel, \& Rodr\'\i guez (1998)
sets an upper limit of 100 km s$^{-1}$ for the velocity of the source with respect
to its galactic position. Then, if GRS~1915+105 is a binary ejected during
the supernova event that produced SNR 45.7-0.4, it required about 400,000 years
to get to its present position. This time interval seems to be much larger that
typical lifetimes for detectable supernovae ($\sim$100,000 years) and we consider
unlikely a common origin for both objects.  

\begin{figure}
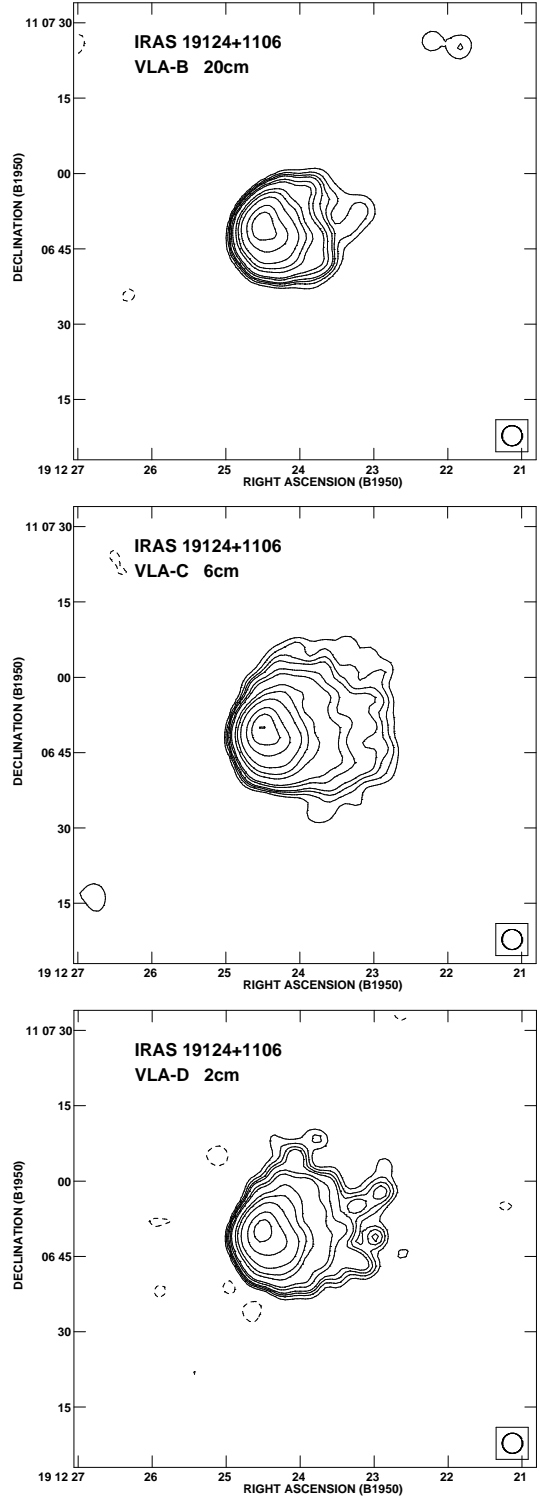

\mbox{}
\vspace{21.0cm}
\includegraphics{ai253.f2a}
\includegraphics{ai253.f2b}
\includegraphics{ai253.f2c}
\caption[]{VLA maps at 20 (top), 6 (middle), and 2 cm (bottom)
of IRAS~19124+1106. The half power contour of the
beam, with diameter of 4$''$, is shown in the bottom
right corner. Contours are -4, 4, 6, 8, 10, 15, 20, 40, 60, 100, 200,
300, and 400 times 0.05 mJy beam$^{-1}$.
}
\label{fig2}
\end{figure}

\begin{figure}
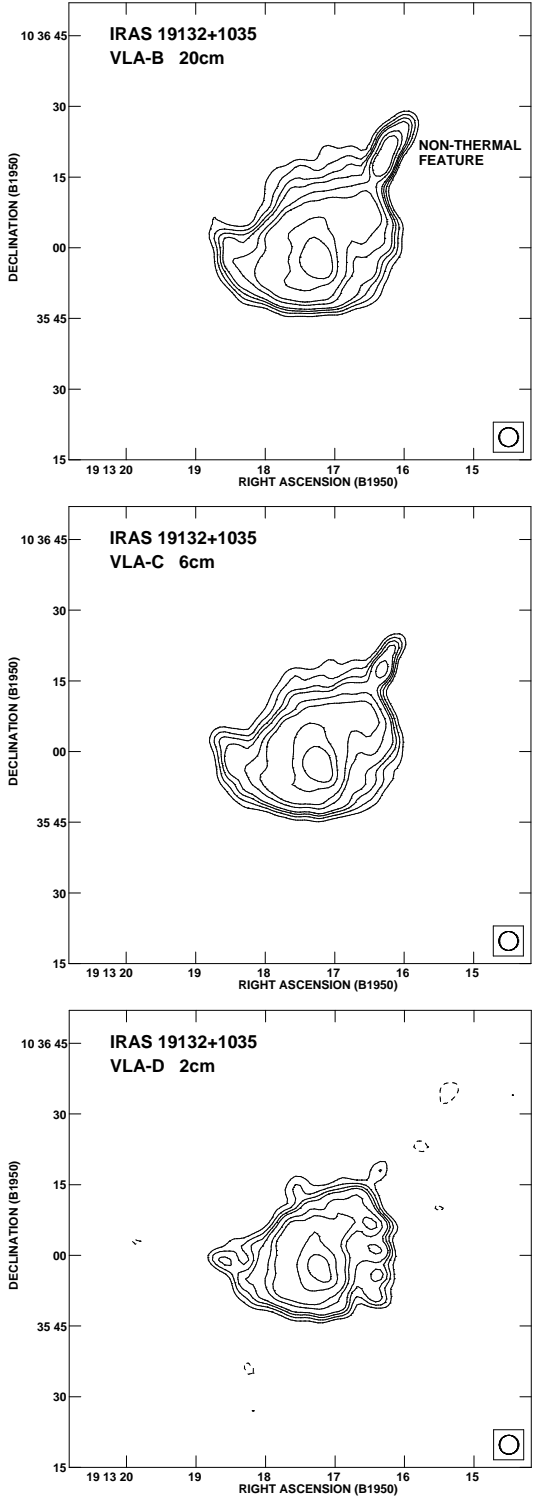

\mbox{}
\vspace{21.0cm}
\includegraphics{ai253.f3a}
\includegraphics{ai253.f3b}
\includegraphics{ai253.f3c}
\caption[]{VLA maps at 20 (top), 6 (middle), and 2 cm (bottom)
of IRAS~19132+1035. The half power contour of the
beam, with diameter of 4$''$, is shown in the bottom
right corner. Contours are -4, 4, 6, 8, 10, 15, 20, 40, 60, 100, 200,
300, and 400 times 0.05 mJy beam$^{-1}$.
}
\label{fig3}
\end{figure}

\section{IRAS~19124+1106 and IRAS~19132+1035} 

These two bright IRAS sources are also relatively bright
radio continuum sources and appear symmetrically
located with respect to GRS~1915+105 (see Figure 1).
Furthermore, their position angle with respect
to GRS~1915+105 is very similar to that of
the relativistic ejecta ($\sim$150$^\circ$).
IRAS~19124+1106 has PSC IRAS flux densities at
12, 25, 60, and 100 $\mu$m of 3.9, 19.6, 260.6, and 581.5 Jy,
while IRAS~19132+1035 has PSC IRAS flux densities at
12, 25, 60, and 100 $\mu$m of 6.9, 34.0, 277.4, and 488.8 Jy.
The IRAS colors are characteristic of embedded H~II regions.

\begin{table}[hbt]
\caption[]{Flux densities of IRAS~19124+1106 and IRAS~19132+1035}
\label{tab:continuum}
\begin{tabular*}{\hsize}{@{\extracolsep{\fill}}lccc}
\hline
& S$_\nu$(20-cm)  & S$_\nu$(6-cm) & S$_\nu$(2-cm) \\
IRAS & (mJy)  & (mJy)  & (mJy)   \\
\hline
    19124+1106  &   114$\pm$4   &    130$\pm$4   &   114$\pm$6   \\
    19132+1035  &   60$\pm$4 &   63$\pm$4 &   52$\pm$6 \\
\hline
\end{tabular*}
\end{table}

\subsection{Matching-beam continuum}

We carried out matching-beam ($\sim$4$''$) 
observations of the two IRAS sources by observing at
20-cm (B configuration; 1997 February 24),
6-cm (C configuration; 1996 April 22), and
2-cm (D configuration; 1996 September 29). In all cases 1923+210 was the phase
calibrator and either 1328+307 or 0134+329 the amplitude calibrator.
Both sources show the flat spectra characteristic of
optically-thin H~II regions (see Table 1).

In Figures 2 and 3 we show the matching-beam maps of IRAS~19124+1106 and IRAS~19132+1035,
respectively,
at 20, 6, and 2-cm. 
IRAS~19124+1106 appears to be a classical cometary H II region.
There is, however, a remarkable feature in IRAS~19132+1035.
To its northwest edge
a linear feature of non-thermal spectrum is clearly observable.
The approximate flux densities for this feature at 20, 6, and 2-cm
are 5, 2, and $\leq$1 mJy, respectively, for a spectral index
of $-$0.8.
This feature points back approximately to GRS~1915+105.
There are several possible explanations for this non-thermal feature.

1) It could be a non-thermal jet produced by the interaction of
the ejecta from GRS~1915+105 with the H~II region.
Furthermore, it can be speculated that the interaction of the
relativistic ejecta with the molecular cloud at this
position (Chaty et al. 1998) could have induced star formation.

2) It could be a background source that happens to lie along the
line of sight. This possibility seems unlikely, given that the
lineal feature is relatively bright ($\sim$5 mJy at 20-cm) and that it is
aligned toward GRS~1915+105. 

3) The feature could be a non-thermal jet emanating from the star
that ionizes the H~II region or even from one of the lower-mass stars that
probably formed in this region, since stars tend to form in groups. 
Radio continuum jets have been observed to emanate from many young stars.
However, the majority are of thermal (i. e. free-free) nature
(Rodr\'\i guez 1997), although a few appear to be non-thermal
(i. e. synchrotron) emitters (Wilner et al. 1997).

\subsection{H92$\alpha$ radio recombination line}

In addition to the continuum study, we observed the H92$\alpha$
recombination line during 1998 January 15.
At that epoch the array was in its lowest angular resolution D
configuration.
The observations were made using 0134+329 as  
absolute amplitude calibrator, 1226+023 as bandpass calibrator,
and 1923+210 as phase calibrator. 

\begin{table*}[hbt]
\caption[]{H92$\alpha$ Radio Recombination Line Observations}
\label{tab:line}
\begin{tabular*}{\hsize}{@{\extracolsep{\fill}}lccccc}
\hline
 & $S_C$ & $S_L$ & $\Delta v$  &  v$_{LSR}$ & Angular \\
IRAS & (mJy) & (mJy) & (km~s$^{-1}$) & (km~s$^{-1}$) & Size ($''$) \\
\hline
19124+1106 & 131.7$\pm$0.9 & 14.9$\pm$1.0 & 25.4$\pm$1.9 & 57.3$\pm$0.9 & 15$\pm$2   \\
19132+1035 & 61.0$\pm$0.3 & 8.9$\pm$1.1 & 23.4$\pm$3.2 & 75.7$\pm$1.6 & 20$\pm$3  \\
\hline
\end{tabular*}
\end{table*}

\begin{table*}[hbt]
\caption[]{Derived Parameters from the H92$\alpha$ and 3.6-cm Continuum Observations}
\label{tab:line2}
\begin{tabular*}{\hsize}{@{\extracolsep{\fill}}lcccccccc}
\hline
Source  &  Distance & $T_e^*$ &  Physical & $\dot N_i$ & ZAMS Star & $n_e$ & M$_{H II}$ & $\tau_c$ \\
 &  (kpc) & (K) &  Size (pc) & (phot s$^{-1}$)& Required & (cm$^{-3}$) & (M$_\odot$) &  (nepers) \\
\hline
IRAS~19124+1106 & 7.4 & 6800$\pm$600 &  0.54 & 6.0$\times$10$^{47}$  & O9 & 5.4$\times$10$^{2}$ & 3.5 & 0.002 \\
IRAS~19132+1035 & 6.0 & 5900$\pm$1000 & 0.59 & 1.7$\times$10$^{47}$  & B0 & 2.6$\times$10$^{2}$ & 2.1 & 0.001 \\
\hline
\end{tabular*}
\end{table*}

\begin{figure}
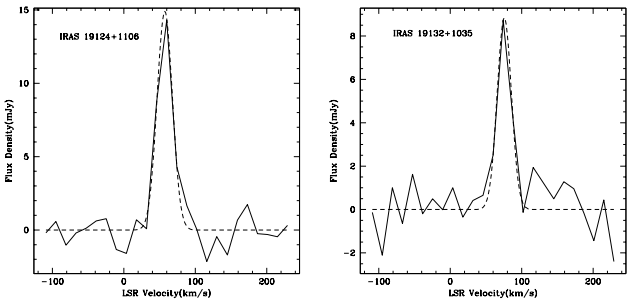

\mbox{}
\vspace{3.9cm}
\includegraphics{ai253.f4a}
\includegraphics{ai253.f4b}
     \caption{H92$\alpha$ spectra of IRAS~10124+1106 (left)
and IRAS~19132+1035 (right).
The dashed line is the least-squares Gaussian fit
to the data.
}
         \label{Fig4}%
    \end{figure}

The spatially-integrated continuum and H92$\alpha$ 
line parameters are given in Table 2
and the H92$\alpha$ spectra are shown in Figure 4.
The observed recombination line parameters are
also typical of H~II regions.
We have used the measured $v_{LSR}$ velocities and the 
galactic rotation curve of Brand \& Blitz (1993)
to estimate distances of 7.4 and 6.0 kpc for IRAS~19124+1106 and IRAS~19132+1035,
respectively. For IRAS~19124+1106 we have a near/far ambiguity
in the kinematic distance and we adopted the far distance
because of its probable association with the G45.46$+$0.06 H~II region complex that
is known to be located behind the tangencial point (Downes et al. 1980).
Given its relatively high v$_{LSR}$, we assumed that IRAS~19132+1035
is located close to the tangencial point.
The derived parameters from the the H92$\alpha$ and 3.6-cm continuum
observations are given in Table 3.

The derived IRAS luminosities are 3.3$\times$10$^4$ $L_\odot$ and
2.4$\times$10$^4$ $L_\odot$, for IRAS~19124+1106 and IRAS~19132+1035,
respectively. These luminosities correspond to an 
O9.5 ZAMS star and a B0 ZAMS star, and agree well with the 
stellar classes inferred from the ionized gas (see Table 3).

We conclude that the parameters of IRAS~19124+1106 and IRAS~19132+1035
are perfectly consistent with those of compact H~II regions
that seem to be closer than GRS~1915+105 (7.4 kpc and 6.0 kpc, instead of 12.5 kpc).
The only anomaly remains the linear non-thermal feature observed in
IRAS~19132+1035. It should also be noted that IRAS~19132+1035
has a sharp edge towards the south, that could be related to
either a bow shock or an ionization front.

\section{Conclusions}

   \begin{enumerate}
      \item We find that the supernova SNR 45.7-0.4 is
too far from GRS~1915+105 too ascribe a common origin to both
objects.

      \item The bright IRAS/radio sources IRAS~19124+1106 and IRAS~19132+1035
appear to be H~II regions ionized by late O or early B stars.
We find that a peculiar non thermal feature is associated with
IRAS~19132+1035, but cannot reach a firm conclusion on
its nature.
   \end{enumerate}

\begin{acknowledgements}
      L. F. R. acknowledges the support of DGAPA, UNAM
and CONACyT, M\'exico.
\end{acknowledgements}

\end{document}